\def\sqig{$\sim\,$} \def\etal{et\,al.} \def\msun{$M_{\scriptstyle\odot}$} 
\def\up#1{$^{\mbox{{\scriptsize #1}}}$} \def\pten#1{$\times10^{#1}$}
\def\deg{$^{\circ}$}  \def\mdot{$\dot{M}$}
\def\Hb{H$\beta$} \def\HeI{He\,{\sc i}} \def\HeIIl{He\,{\sc ii}\,$\lambda$4686} 
\def\sqiggt{\hbox{\rlap{\lower.55ex \hbox {$\sim$}}
\kern-.3em \raise.4ex \hbox{$>$}\,}}
\documentstyle{mn}
\begin{document}
\title[SW~Sextantis stars]{The phase 0.5 absorption in SW~Sextantis-type cataclysmic
variables}
\author[C. Hellier]{Coel Hellier \\
Department of Physics, Keele University, Keele, Staffordshire, ST5 5BG,
U.K.}
\date{Accepted for PASP April 1998 issue (despite the MN macros)}
\maketitle
\begin{abstract}
The SW~Sextantis stars are a group of cataclysmic variables with 
distinctive observational characteristics, including absorption
features in the emission line cores at phases 0.2--0.6.
Hellier and Robinson have proposed that these features are caused by
the accretion stream flowing over the accretion disk. However, in a
simple model the absorption occurred at all orbital phases, which
is contradicted by the data. I show that invoking a flared accretion
disk resolves this problem.
\end{abstract}

\section{Introduction}
The SW~Sextantis stars are a subclass of the novalike variables, which are
themselves a subclass of the cataclysmic variables (CVs). The qualification for
novalike status is a stable accretion disk, presumably due to a mass transfer
rate sufficiently high to prevent the instabilities associated with dwarf novae
(e.g.~Warner 1995). The SW Sex stars are novalikes showing all or most of the
following properties (e.g.~Thorstensen \etal\ 1991): (i) single-peaked emission
lines, particularly \HeIIl, incompatible with an origin in a Keplerian disk;
(ii) gross asymmetries in the emission lines from the disk, so that they do not
reflect the orbital motion of the white dwarf, (iii) peculiar `phase 0.5'
absorption features in the core of the line during the orbital phases
\sqig0.2--0.6; (iv) a tendency to have orbital periods in the range 3--4 hr,
just above the CV period gap; and (v) a high probability of being eclipsing
(this one is most likely a selection effect).

Many ideas have been proposed to explain SW~Sex stars, the favourites being
strong accretion disk winds (Honeycutt \etal\ 1986; Dhillon \etal\ 1991; 
Hellier 1996); magnetically controlled accretion (Williams 1989; 
Casares \etal\ 1996); and mass-transfer streams which penetrate or flow over 
the disk (Shafter \etal\ 1988; Hellier and Robinson 1994).

This paper addresses the combination of those ideas proposed in Hellier and
Robinson (1994; hereafter Paper~1) and Hellier (1996; Paper~2), although 
for alternative views and discussion of other models see Dhillon \etal\
(1991), Casares \etal\ (1997) and Hoard and Szkody (1997). We suggested that
SW~Sex stars are novalikes with abnormally high mass-transfer rates. This
causes, first, strong winds from the inner disk or boundary layer,
explaining the single-peaked line profiles. Second, it allows the accretion
stream to overflow the initial impact with the edge of the disk rim, and
continue on a free-fall trajectory to a second impact much nearer the white
dwarf. Line emission from this re-impact causes the highly asymmetric line
profiles. Further, we proposed that the stream is seen in absorption between
the initial impact with the disk and its re-impact. This means that the
normal emission `S-wave' common in other CVs is absent, or even in
absorption. Also, we noted that the absorption from the stream could 
explain the phase 0.5 absorption features of SW~Sex stars. 

The remaining characteristic, the concentration on orbital periods just above
the period gap, presumably results from evolutionary effects driving systems
with those periods to higher mass transfer rates (\mdot; e.g.~Shafter 1992).
Indeed, in Paper~2 we noted that SW~Sex stars often show VY~Scl low states,
implying that the abnormally high \mdot\ in the high state is balanced by
periods of much lower \mdot. This could result from irradiation-driven
feedback cycles in \mdot\ (e.g.~Wu \etal\ 1995).

In Papers 1 and 2 we computed the velocities expected from an accretion stream
flowing over the disk, as a function of orbital phase, and turned these into
model line profiles for comparison with data from PX~And and V1315~Aql. Overall
the simulations supported the model, showing that the overflowing stream had
the right velocity variations to explain the distorted line profiles and the
phase 0.5 absorption. However, these simulations had a fundamental limitation
in that they calculated velocities only, making no allowance for obscuration of
one component by another, and so could not reproduce variations in the strength
of features round the orbital cycle. Thus the simulations contained `phase 0.5
absorption' at all orbital phases, whereas the data doesn't.

Since several authors (e.g.~Casares \etal\ 1996; Hoard and Szkody 1997)  have
cited this flaw as a primary reason for doubting the model,  I have re-written
the simulation code to include all spatial  and obscuration effects (it is
still a simple geometric model, though, making no attempt at radiative
transfer). I pick up a suggestion from Paper~1 that disk flaring can confine
the absorption to a limited phase range. This gives a much closer resemblance
to the data, solving the biggest discrepancy between the data and models from
Papers 1 and 2.

\section{The modelling code}
The modelling code is a development of that presented in Papers 1 and 2.
It adds together the spectrum expected from a flared Keplerian disk, a 
single-peaked profile expected from a wind, and a component with the 
velocities of a stream flowing over the disk. In contrast to the previous
versions of the code it calculates spatial obscuration including: eclipses
of the disk and stream by the secondary; obscuration of parts of the disk
by the stream; the degree to which the stream can be seen over the rim
of the flared disk; and obscuration of the stream by other regions of 
stream.

The code employs a large number of parameters, although many of them
are either confined to a narrow range by other data, or have little
effect on the simulation. In what follows I give further details of the
code and justification for the parameters adopted.

I used an orbital period of 3.5 hr, primary and secondary masses of 0.8 and
0.3 \msun\ and a disk radius of 4\pten{8}\,m, typical of SW~Sex systems.
Changes in these parameters tend to rescale the simulated spectra, rather than
changing details of the absorption, which is the primary interest of this
paper. 

Obscuration effects in these deeply eclipsing binaries are very sensitive to 
the inclination and the disk flaring angle. I used an inclination, $i$, of 
82\deg, from Dhillon \etal's (1991) estimate of 82.1\,$\pm$\,3.6 for
V1315~Aql. For the flare (opening semi-angle, $\alpha$) I used 4\deg. This is
based on  theoretical estimates that the photosphere has an $\alpha$ of
3--4\deg\ for a high \mdot\ system (Smak 1992; Wade 1996), and on two
observational estimates: In Z~Cha during outburst Robinson \etal\ (1995)
measured $\alpha$ = 8\deg\ (although consideration of limb-darkening reduces
the estimate to 4\deg; Wood, private communication). Also, an $\alpha$
$>$3.5\,$\pm$\,1.6\deg\ in DQ~Her results from an inclination of
86.5\,$\pm$\,1.6\deg\ (Horne \etal\ 1993), and the fact the no X-rays
are seen, so that the white dwarf is presumably always obscured.

We expect that the inner disk produces line absorption and the outer disk
line emission. The code includes this effect, giving a rough match to the
empirical data presented by Rutten \etal\ (1993) for the novalike UX~UMa 
(the detailed implementation of this made very little difference to the 
simulation). The code doesn't include continuum or line flux from the outer
wall of the flared disk, assuming this region to be a minor contributor 
compared to the irradiated surface of the disk. 

I also included a single-peaked profile, moving with the velocity of the white
dwarf, based on the synthetic wind profile of Hoare (1994). In the Balmer lines
this component simply fills in the double-peaks of the disk profile, but it
dominates in lines such as \HeIIl\ (Dhillon \etal\ 1991; Paper~2). This
component is often attributed to an accretion disk wind; however even if this
interpretation is incorrect it is an empirical fact that SW~Sex stars have this
component, and that in the Balmer lines it is largely un-eclipsed. Thus, in my
code the wind component is not eclipsed. Since the simulation produces a plot
normalised to the continuum (to match the presentation of the data in Papers 1
and 2) this means that the wind is enhanced during eclipse.

For the stream flowing over the disk I used the calculations of Lubow (1989).
This assumes that the stream is wider than the disk at initial impact so that
some portion of it continues on a ballistic trajectory to a second impact
near the white dwarf (see Papers 1 and 2 for a fuller account). The 
flare angle of 4\deg\ results in a disk height similar to that of the stream
as calculated by Lubow (1989), which is \sqig0.13\pten{8}\,m for typical
SW~Sex parameters. This presents a potential problem for the
disk-overflow hypothesis, in that too thick a disk will block the stream.
Note, though, that the height relevant to the optical properties is that of the
photosphere, and that material at that height is not necessarily 
substantial enough to impede the stream. Using hydrodynamical simulations rather
than the analytical treatment of Lubow (1989), Armitage and Livio
(1996; 1997)
obtain, unsurprisingly, a far messier result. They confirm that some of the 
stream material overflows the disk, but in a wider fan beam and with a 
trajectory deflected from the original direction. Despite this, I use the
Lubow (1989) calculations for the height and direction of the stream, simply
because they are much easier to compute. Note that my results are
fairly insensitive to the overflow trajectory, but if future
simulations produce outcomes very different from the ballistic approximation
then they could refute my model.

As in Papers 1 and 2\,\ I suppose that for the first two-thirds of the stream
trajectory over the disk the stream is seen in absorption. Then, for the
final third when it re-impacts the disk, it produces line emission.
The absorption profile of the stream was taken (somewhat arbitrarily) to
be the same depth as that of the inner disk (20\%\ of the continuum). In fact,
to a very large extent the depth of the absorption from each stream element
can be traded against the width of the stream. I therefore used a fixed 
absorption depth and let the stream width be a free parameter. To
recapitulate, the effect of the stream in the code is to block the 
contribution from any disk element which is behind the stream, and to add
in absorption for any stream element in its first two-thirds and emission
for elements in the last third.

In constructing the synthetic line profiles each pixel of each component
contributed a Gaussian profile centered on the correct line-of-sight
velocity with a width ($\sigma$) of 25\%\ of the speed of the element.

\begin{figure}\vspace*{7.2cm}
\caption{Typical \Hb\ line profiles of V1315~Aql showing the absorption in
the line core. The spectra, from the dataset presented in Paper~2, were taken
at orbital phase 0.48 (upper) and 0.55 (lower). The lower curve was displaced
downwards by 50 counts, for clarity.}
\includegraphics{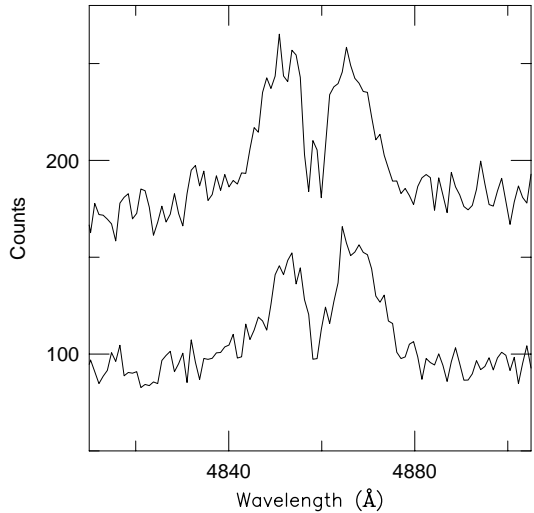}
\end{figure}

\section{Results}
To show the effect I am trying to model, Fig.~1 presents typical \Hb\ line
profiles cut into by the phase 0.5 absorption in the line core. I also show,
Fig.~2, the velocity trend of the absorption and the range of phases over which
it appears. These measurements are reproduced from Thorstensen \etal\ (1991).
They are from a metal complex at $\lambda5175$ in PX~And, chosen because it
shows no apparent emission and so allows the cleanest measurement of the
absorption. Thorstensen \etal\ noted that they obtained similar results from
the \HeI\ lines, but these were less reliable due to the strong and variable
emission. Further presentations of data can be found in Papers 1 and 2 and
references therein.

To compare with the data, Fig.~3 shows my simulation of the Balmer line of an 
SW~Sex star as a function of orbital phase. The central core of
the line is made up of the disk and wind components. The wind is uneclipsed,
leading to the significant increase in equivalent width during eclipse (phase
1). A high amplitude S-wave, arising from the stream re-impact with the disk,
has maximum blueshift at phase 1 and maximum redshift at phase 0.5. The
absorption, caused by the stream flowing over the disk, appears on the red side
of the line at phase \sqig0.2 and moves to the blue, before disappearing at
phase \sqig0.7.

\begin{figure}\vspace*{9.7cm}
\caption{The velocity variation of the $\lambda$5175 absorption complex from
Thorstensen \etal\ (1991). The circles and triangles were cross-correlation
and eye estimates respectively. The symbols at phase 0.66 aren't filled,
indicating that these points are less reliable since the absorption was very
weak. The phase range and velocity trend should be compared with the 
absorption in Fig.~3. }
\includegraphics{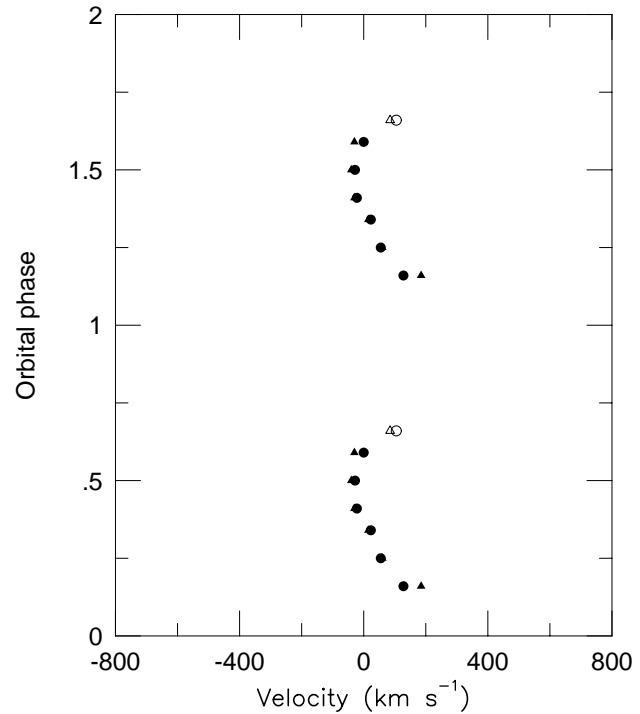}
\end{figure}

The comparison of simulation to data shows an excellent match to the phases at
which absorption occurs (0.2--0.7) and to the velocity trend during those
phases. The match is not so good to the $\gamma$ velocity of the absorption, 
although Thorstensen \etal\ (1991) note that the $\lambda5175$ feature is
probably a blend of several lines of uncertain relative contribution, so the
$\gamma$ velocity of the data is essentially arbitrary. More problematical is
that the velocity change has a bigger amplitude in the model than the data. 
This could result from my use of free-fall velocities. The simulations
by Armitage \&\ Livio (1997) show that the velocity of the overflowing
material is reduced by up to a half by interactions with the disk, although
the reduction depends to a large extent on whether an isothermal or an
adiabatic equation of state is used. This might solve another outstanding
issue for the disk-overflow model, since when modelling V1315~Aql (Paper 2)
I found that the best match to the data occurred with velocities 25\%\ below
the free-fall velocity, which is now explained by Armitage and Livio's (1997) work.
This is one reason why tomograms of SW~Sex stars often show most
line emission in the lower-left quadrant, at velocities lower than
that of the ballistic stream. The other reason is that such bright regions
on a tomogram can be caused by the overlap of emission from
the overflow and disk components, and do not necessarily indicate
 components in their own right (see Papers~1 \&\ 2).

To summarise, a flared disk solves the  problem of confining the absorption to orbital
phases around 0.5. The only parameter having a significant effect on this
aspect of the simulation is the flare angle $\alpha$, or more precisely the 
departure of $i\,+\,\alpha$ from 90\deg. If $i\,+\alpha$ is \sqiggt 10\deg\
the absorption becomes weaker and visible at all orbital phases, as in the
simulations in Papers 1 and 2.

The stream width used in the simulation was 0.4\pten{8}\,m,
which is 1/10\up{th}\ of the disk radius. The resulting absorption depth is
comparable to that in SW~Sex stars (e.g.~Fig.~1). Again, the absorption depth
from each stream element (currently 20\%) could be altered to give the
same result for a different stream width. Note, also, that the simulated
absorption could be filled in at the extremes of its appearance (phases 0.2 and
0.6) by increasing the relative intensity of the emission from the re-impact.
This would produce absorption at a narrower range of phases, centered at
0.4--0.5, as is more common in the Balmer lines of these stars.

\begin{figure*}
\includegraphics{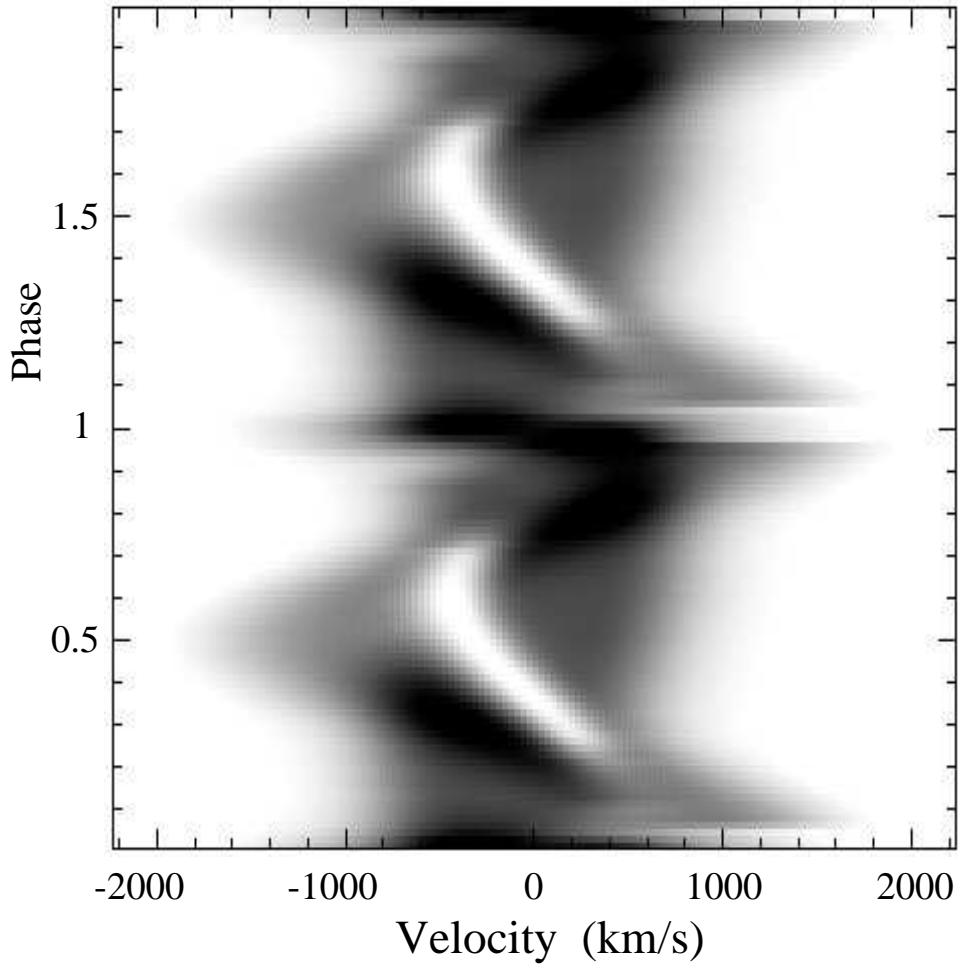}
\vspace*{14cm}
\caption{The simulation of the Balmer line of an SW~Sex star, using the 
modelling code described in Section~2. The `phase 0.5 absorption' moves
from the red wing at phase 0.2 to the blue wing at phase 0.6. Darker
coloring implies greater emission.}
\end{figure*}

\section{Conclusions}
The `phase 0.5 absorption' features seen in SW~Sex stars can be explained by a
proportion of the accretion stream flowing over the disk. The fact that
absorption is seen only at phases 0.2--0.6, and not at every phase as
predicted by simulations in Papers 1 and 2, is explained if the accretion disk
is flared at an angle of $\approx$ 4\deg. The simulation code has been
re-written to include all spatial and obscuration effects, and the results in
Papers 1 and 2 are still valid. If disk-overflow is a generally
correct model for SW~Sex stars the degree of overflow could still
vary with time or between systems. A reduction in disk-overflow
would leave an SW~Sex star looking much like a normal novalike
(as seems to have been the case for SW~Sex when observed by Dhillon, Marsh 
\&\ Jones 1997). In a high inclination novalike the splash caused
by the stream--disk impact can itself produce absorption when it
is at inferior conjunction, which would produce absorption dips
centered on phase \sqig 0.8, rather than the phase 0.2--0.6 characteristic
of the SW~Sex stars. Dips at phase \sqig 0.8 are seen in novalikes
such as TV~Col (Hellier 1993) and BP Lyn (Hoard \&\ Szkody 1997).

\end{document}